# Investigation of thickness dependent composition of boron carbide thin films by resonant soft x-ray reflectivity

P.N.Rao[a,b,*], R.K.Gupta[a], K. Saravanan[c], A. Bose[d], S.C. Joshi[d], T.Ganguli[a,b], S.K.Rai[a,b]

[a]Synchrotrons Utilization Section, Raja Ramanna Centre for Advanced Technology, Indore 452013, India

[b]Homi Bhabha National Institute, Anushakti Nagar, Mumbai 400 094, India

[c]Materials Science Group, Indira Gandhi Centre for Atomic Research, Kalpakkam - 603102, Tamilnadu, India

[d]Proton Linac and Superconducting Cavities Division, Raja Ramanna Centre for Advanced Technology, Indore 452013, India

**Abstract:** Boron carbide thin films of different thicknesses deposited by ion beam sputtering were studied. The deposited films were characterized by grazing incidence hard x-ray reflectivity (GIXR), resonant soft x-ray reflectivity (RSXR), x-ray photo electron spectroscopy (XPS), resonant Rutherford backscattering spectrometry (RRBS), and time of flight secondary ion mass spectrometry (TOF-SIMS). An in-depth profile of the chemical elements constitute the films is reconstructed based on analysis of reflectivity curves measured in the vicinity of B K-edge. The composition of films is closely dependent on film thickness. Boron to Carbon (B/C) ratio reaches to ~4 as the thickness of deposited films increases. The B/C ratio estimated from RSXR measurements are in agreement with the RRBS measurements. TOF-SIMS data also suggested that decrease in boron content with decrease in film thickness. XPS measurements confirm the presence of little amount of B atoms on the surface of low thickness film.

Key words: Boron carbide, optical index, composition, protecting layer, X-ray multilayers, resonant reflectivity

[*]Author to whom correspondence should be addressed.    Electronic mail: pnrao@rrcat.gov.in

## 1. Introduction

Boron carbide is an important x-ray optical element in both hard and soft x-ray regions [1-4]. It is also an important barrier material to minimize the inter diffusion in multilayers (MLs) [5]. Boron carbide which has a very high melting and sublimation point is one of the suitable candidates for free electron laser applications [6-8]. It is used as capping layer on top of the ML structure to protect ML structure from oxidation [9]. Boron carbide is also a promising material for the next generation photo lithography applications at 6.x nm (the value of x still has to be determined by industry) wavelength [10, 11]. The compositional changes in the boron carbide causes significant changes in its optical constants in the vicinity of B K-edge and limits maximum achievable throughput from boron carbide based MLs [12]. The optical band gap of boron carbide and the properties of boron carbide/Si diodes can be varied by simply changing composition of boron carbide [13-15]. In many practical applications boron carbide thin films with thickness ranging from a fraction of a nanometer to several nanometers have been used. It is thus important to study the thickness dependent compositional changes in boron carbide thin films.

In our previous work [12, 16] on $C/B_4C$ MLs, we observed deviation in derived optical constants of boron carbide from the values available in literature [17]. A detailed analysis in Ref.12, suggested that the presence of excess carbon into the boron carbide layer. In that work we concluded that presence of excess carbon in the boron carbide layer is either due to inter-diffusion of C atoms from the C layer or due to change in stoichiometry of deposited boron carbide layer from the target ($B_4C$) stoichiometry. To investigate further the compositional changes in boron carbide we have chosen thin films of boron carbide in this study. In general ion and electron beam techniques are used to determine concentration profiles of thin films. Electron beam techniques like X-ray photo electron spectroscopy and Auger electron spectroscopy are surface sensitive and in-depth information is obtained by etching out the material. These techniques are destructive in nature and consequently repetitive measurements on specimen cannot be possible [18]. Grazing incidence hard x-ray reflectivity (GIXR) is routinely used technique to derive layer thicknesses, surface and interface roughness of thin films and MLs. However, the availability of synchrotron sources with high brilliance and tunable energy provide the opportunity for researchers to explore additional properties apart from structure using reflectivity technique. In the vicinity of the absorption edges optical index is strongly depends on the composition of layers. The reflectivity measured in the vicinity of absorption edges has opened up possibility to derive composition of thin films in nondestructive way. The angular dependence inherent to reflectivity measurements provides the depth resolution down to nm with penetration depths over hundreds of nanometers. Resonant soft x-ray reflectivity (RSXR) has been used for characterization in low contrast organic thin films [19] and compositional analysis in thin film [20] and periodic multilayers [12].

In this paper we investigate the compositional changes in boron carbide thin films as a function of its thickness by combining GIXR and RSXR. X-ray photo electron spectroscopy (XPS), resonance Rutherford backscattering spectrometry (RRBS) and time of flight secondary ion mass spectrometry (TOF-SIMS) measurements were performed to complement the observation made from RSXR.

## 2. Theoretical background

In the x-ray region, response of the medium is described by energy dependent complex refractive index $n=1- \delta + i\beta$, where $\delta$ (dispersion coefficient) and $\beta$ (absorption coefficient) are the optical constants [1]. The reflection or refraction at a boundary between two media occurs due to change in refractive index. Specular x-ray reflectivity is one of many x-ray scattering techniques used to probe the profile of film along the direction perpendicular to the sample surface. When probing thin films with reflectivity, one generally characterizes the interference of rays reflected from different interfaces as a function of momentum transfer vector q, defined as $q = q_{in} - q_{out} = 4\pi \sin\theta/\lambda$, where $q_{in}$ and $q_{out}$ are incident and reflected momentum transfer vectors respectively, $\theta$ is the grazing incidence angle and $\lambda$ is the wavelength of probing beam. The modulations in reflectivity profile measured using shorter wavelengths were analyzed to get the geometrical parameters like thickness and root mean square (rms) interface and surface roughness. In case a film contains low-Z elements, the difference in electron density among low-Z elements is small, thereby limiting the observable contrast in conventional GIXR. We thus can get limited information on the depth distribution of film contains low-Z elements using GIXR.

In the vicinity of the absorption edge both the optical constants δ and β undergo strong variations. For many light elements absorption edge falls in the extreme ultraviolet/soft x-ray region. Any small changes in the composition of film gives rise to a significant change in optical index contrast and that can give a clear contrast in a reflectivity profile. Thus analysis of the reflectivity curves measured near the absorption edges gives the concentration profile of all elements in the sample. The angular dependence inherent to reflectivity measurements brings furthermore the spatial selectivity. If a sample contains several chemical elements the optical indices can be represented as

$$\delta = 2.7007\times 10^{-4} \lambda^2 \rho \frac{\sum_j X_j \left(f^0_{NR,j} + f'_{R,j}(\lambda)\right)}{\sum_j X_j \mu_j} \qquad (1)$$

$$\beta = 2.7007 \times 10^{-4} \lambda^2 \rho \frac{\sum_j X_j f_{R,j}^{"}(\lambda)}{\sum_j X_j \mu_j} \quad (2)$$

where $\lambda$ is the incident wavelength [nm], $\rho$ is the density [g/cm$^3$], $X_j$ is the atomic fraction of j atoms, and $\mu_j$ is the atomic weight of j atoms [g/mol], $f_{NR}^o(q)$ is the non-resonant atomic scattering factor (ASF), $f_R^{'}(\lambda)$ and $f_R^{"}(\lambda)$ are the dispersion and absorption corrections to the ASF arise from the bounded electrons in an atom. In the present study we deduce the concentration profiles of all chemical elements present in the film from the optical index derived from the reflectivity curves measured in the vicinity of B K-edge. In order to take into account the effects of bonding between B and C in B$_4$C on the ASFs, we used ASFs for B derived from magnetron sputter deposited boron carbide thin films [17]. ASFs for non-resonating atoms like C and O were taken from the Henke *et al*. [21] tabulated values. The mass density obtained from GIXR measurements were used and best fits to the derived optical constants was obtained by varying the atomic fraction of different chemical elements in the film.

### 3. Experimental techniques

*3.1. Sample preparation*

Boron carbide thin films of various thicknesses were deposited on ultrasonically cleaned Si (100) wafer using ion beam sputtering technique. We used commercially available four inch sputtering target of 99.5% purity for B$_4$C. Ar ions were used to sputter the target materials. Target material was pre sputtered for 30 min before the deposition of films. In the present study, four boron carbide thin films of thickness, (d) ~ 10 nm, 30 nm, 40 nm and 80 nm have been used. These samples were named as S1 (d~10 nm), S2 (d~30 nm), S3 (d~40 nm), and S4 (d~80 nm).

*3.2. Reflectivity measurements*

GIXR measurements were carried out [Bruker discover D8 diffractometer] using Cu $K_\alpha$ radiation ($\lambda$=0.154 nm). The angle dependent soft x-ray reflectance measurements in the vicinity of B K-edge were carried out using the reflectivity beamline at Indus-1 synchrotron facility. This beamline, having a toroidal grating monochromator, delivers photons in the range of 4-100 nm, with high flux (~$10^{11}$ photons/second), and has a moderate spectral resolving power ($\lambda/\Delta\lambda$) of 200-500. Various absorption edge filters are provided in the beamline to suppress the higher order contamination from the monochromator. A nonlinear least square curve fitting technique based on $\chi^2$ minimization method was applied for the determination of microstructural parameters and optical constants from reflectivity curves [22].

### 3.3. Resonance Rutherford backscattering spectrometry (RRBS)

In general Rutherford backscattering spectrometry (RBS) is used for analysis of heavier elements on lighter substrate. However RBS is ineffective for detection of lighter elements on heavier substrate due to small scattering cross-section and overlap of background signal coming from heavier substrate. Resonant Rutherford backscattering spectrometry (RRBS) has been widely used for analysis of light elements on heavier substrate. In order to quantification of the elements oxygen, boron and carbon presents in the sample S4, RRBS measurements have been performed. The RRBS measurements were performed using 3.045 MeV $He^{++}$ particles for the quantification of oxygen, 3.9 MeV $He^{++}$ particles for the quantification of boron and 4.27 MeV $He^{++}$ particles for the quantification of carbon as alpha particles has enhanced scattering cross-section with these elements at these energies. The 1.7 MV Tandetron accelerator (HVEE, The Netherlands) available at IGCAR, Kalpakkam was used for the RRBS analysis. The backscattered particles were detected using a Si surface barrier detector kept at an angle of 165° with respect to the incident beam direction. The concentrations of oxygen, boron and carbon in the samples were obtained from the best-fit of RRBS data using the SIMNRA program [23].

*3.4. X-ray photo electron spectrometry (XPS)*

XPS measurements were carried out using photoelectron spectrometer at a base pressure better than $5 \times 10^{-10}$ mbar. Al $K_\alpha$ radiation ($\lambda$=8.34 nm) was employed for recording the spectra with the source operated at an emission current of ~10 mA and an anode voltage of ~ 10 kV. Samples were in-situ cleaned by light Ar ion sputtering before measurement.

*3.5. Time of flight secondary ion mass spectrometry (TOF-SIMS)*

TOF-SIMS depth profile analysis of boron carbide thin films were carried out in interlaced mode with pulsed $Bi_1^+$ at 30 keV (~5pA) as the analysis gun and 1keV $Cs^+$ as the sputter gun operating at a constant current of ~75nA. The area of analysis across all measurements was 100µm x 100µm inside the sputter area of 300µm x 300µm. The sputter gun parameters were selected to ensure that erosion is slow enough to reveal the variation in low thickness boron carbide thin film sample which was again used to profile the thicker sample.

## 4. Results and discussion

*4.1. GIXR measurements*

Reflectivity curves obtained from the GIXR measurements at Cu $K_\alpha$ radiation ($\lambda$=0.154 nm), along with the theoretically calculated curves as a function of scattering vector ($q_z$) are shown in Fig. 1. The oscillations are due to interference between the rays reflected from different interfaces in the sample and the damping of these oscillations depends on the surface roughness of the sample. From Fig. 1, it is clear that the different samples have different $q_z$ range. This indicates that the root mean square (rms) surface roughness of the grown film changes significantly with thickness and deviates from the substrate rms roughness. We assumed two layer model consisting of boron carbide/SiO$_2$/Si to simulate the measured GIXR curves. In general, an oxide layer is formed on top of Si substrate due to exposes to ambient. The difference between measured and calculated reflectivity curves is minimized by changing thickness, density and roughness of the assumed layers. The coated film with uniform density gives a satisfactory fit for all the samples except for the sample S2. In fact, this coated layer needs to be divided in two parts of different densities to obtain a best fit for the sample S2. Therefore, we use a tri-layer model consisting of boron carbide layer-2/boron carbide layer-1/SiO$_2$/Si. The best fitted model suggested that the density of coated layer-1 is less than the coated layer-2. Thickness, density and roughness of deposited films obtained from the best fitted model shown in Fig. 1 are presented in Table 1. From Table 1, it is clear that the rms roughness of the films increases with increasing the thickness. As the thickness of the films reduces the density of the film reduces and deviated significantly from the bulk value (~2.52 g/c.c). The geometrical parameters of deposited films were known. Using GIXR we cannot distinguish between B and C atoms because of low electron density difference (~2%). The compositional profile of film consisting of low electron density contrast elements cannot be determined using GIXR. The compositional profile of deposited layers can be determined by simulating the measured soft x-ray reflectivity curves in the vicinity of the B K-edge. The geometrical parameters like thickness and rms roughness obtained from analysis of GIXR curves were used as input parameters while analysis of RSXR data.

*4.2. RSXR measurements*

In the vicinity of the absorption edge both the optical constants δ and β underwent strong variation. RSXR utilizes the optical contrast due to both δ and β and shows strong modulation in reflectivity profile as the energy of incident radiation changes. An in-depth profile of the chemical elements composing the film is reconstructed based on analysis of reflectivity curve measured in the vicinity of the absorption edge. The factors which influence the determination of optical constants and hence a reconstructed composition profile of film using soft x-ray reflectivity are (i) A contaminated layer (CLR) present on top of the sample surface (ii) geometrical parameters like thickness and roughness. The necessity of introducing the CLR and impossibility of an adequate description experimental soft x-ray reflectvities was demonstrated by Filatova *et al*., [24-26] if this layer was neglected. Because of refraction correction in soft x-ray region wrong thickness values leads to different optical constants/composition. The surface and interfacial roughness results in a deformation of the reflectivity curve which increases with increasing grazing angle and hence in an additional error in derived optical constants. A need to combine the reflectivity techniques to deduce the geometrical parameters and to take geometrical parameters into account during the reconstruction of composition of deposited films was explained well in Ref. 25. CLR on the sample surface is mainly consisting of hydrocarbons, water, and oxygen. Using samples of different thicknesses can help us in estimating the CLR uniquely. The geometrical parameters obtained from the GIXR are used as input parameters while analyzing the RSXR data and best fit is obtained by changing the optical constants of the deposited films. This method improves the reliability of reconstructed composition profile of film.

RSXR measurements were performed in the energy ranging from 184 to 197 eV. Fig. 2, shows the measured and simulated RSXR curves of the sample S3 at selected energies in the vicinity of B K-edge. From Fig. 2, it is clear that reflected intensities are entirely different for different energies. In the vicinity of the absorption edge optical constants changes significantly as energy changes and that can gives a clear change in a reflectivity pattern. The derived optical constants are in close agreement with the values reported by R. Sofuli *et al.*, [17] measured on the magnetron sputtered boron carbide thin films of composition 74 at.% B, 20 at.% C and 6 at.% O. The derived optical constants of the sample S4 which are not presented in the present paper are similar to the sample S3. The close agreement in optical constants with the reported values suggested that the samples S3 and S4 have composition similar to magnetron sputtered boron carbide thin films [17]. The presence of oxygen in our grown films is either due to oxygen coming from the sputtered target (i.e was incorporated during target fabrication) or due to oxygen being present in the environment during deposition.

The measured RSXR curves along with simulated curves and in-depth $\delta$ profile obtained from it are shown in Fig. 3(a) and (b) respectively of the sample S2 at selected energies in the vicinity of B K-edge. A shoulder in the reflectivity profile appears at 184.5 eV and disappears for other energies. At lower energies away from the B K-edge of $B_4C$ (~189.6 eV) $\beta$ decreases and $\delta$ become positive results in appearance of shoulder in the reflectivity curve. The modulations in the reflectivity result from the superposition of the reflected signals from different interfaces in the total stack. The simulated curves are obtained by varying optical constants of each constituent layers. From Fig. 3b, it is clear that value of $\delta$ of layer-1 is higher than the layer-2. At energy 189.6 eV, $\delta$ value of layer-1 is positive where as it is negative for layer-2. The positive value of $\delta$ is due to presence of excess C into layer-1. This suggests that there is a gradual increase in B/C ratio with increase in thickness. At this energy B is a resonating atom and C is a non resonating atom. Polarizability of the resonating atom is out of phase with the electric field which results in real part of ASF or $\delta$ become negative. Polarizability of non-resonating atom is in phase with electric field which results in real part of ASF or $\delta$ become positive. Any excess C present into the boron carbide layer results in change in sign reversal of ASF or $\delta$.

Fig. 4(a) and (b) shows optical constants profile of layer-1 and 2 respectively as a function of photon energy in the vicinity of B K-edge as obtained from the RSXR fits. For quantitative analysis, atomic fractions of these two entities namely B and C can be varied to obtain the best fit to the measured optical constants. Oxygen which comes from the target or the deposition environment is also included to determine the composition of the films. In these calculations, the mass density of layers obtained from the GIXR measurements, ASFs of B derived from the measured optical constants reported by R. Sofuli *et al.*, [17] and tabulated [21] ASFs for C and O were used. To obtain the best fit for the measured optical constants the atomic fraction of these three entities were varied. A composition of layer-1 containing 50 at. % B, 45 at. % C, and 5 at. % O gives the best fit to the experimental data. A composition of layer-2 which gives the beat fit to the experimental data is 63 at. % B, 32 at. % C, and 5 at. % O.

Fig. 5 shows the measured and simulated RSXR curves of the sample S1 at selected energies in the vicinity of B K-edge. A shoulder in lower $q_z$ appears and reflected intensities are similar for all energies. From Fig. 5, it is clear that reflected intensities are similar for different energies. An appearance of the shoulder in lower $q_z$ and flat response of the reflectivity profile suggests that no or little amount of B atoms are present in the deposited layer. The derived $\delta$ varies from 0.0048 to 0.0052 and $\beta$ varies from 0.003 to 0.0017 as the energy changes from 196.8 to 184.5 eV. To derive the composition the mass density obtained from GIXR were used. The Composition of layer is 23 at. % B, 62 at. % C, and 15 at. % O results into a good fit to the experimental data.

### 4.3. RRBS measurements

The RRBS spectra recorded with three different resonant energies for the quantification of O, B and C are shown in Fig. 6(a), (b) and (c) respectively. The signals backscattered from O, B, and

C are clearly seen in the spectra. The simulated RRBS spectra are also shown in the Fig. 6 along with the experiment. From the best-fit of RRBS spectra the estimated atomic composition of O, B and C was found to be 8%, 74% and 18% respectively. This corresponds to B/C ratio ~4 which is consistent with the sputtering target stoichiometry. The composition estimated from the derived optical constants from the SXR measurements are also in agreement with the composition estimated from the RRBS.

### 4.4. XPS measurements

In addition to reflectivity, we also performed spectroscopic measurement like XPS, which deals with electronic states of elements. Using this technique we can get elemental and compositional information. XPS spectra were taken on the samples S1 and S2 after $Ar^+$ etching for 5 minutes using ion energy of 1 keV. Fig. 7, shows survey scan of samples S1 and S2. XPS spectrum of sample S2 contains significant contribution from O, C and B, where as in sample S1, signal from B is very weak. These XPS spectra suggested the depletion of B from the surface of the sample S1 which corroborates with the observation made from the RSXR measurements. The core level spectra of C 1s, B 1s and O 1s obtained from the surface of sample S1 together with the spectrum deconvolution and Shirley background is shown in Fig. 8. Fig. 8a shows an asymmetric C 1s spectrum where two different chemical states at around 282.8 and 285 eV were identified after deconvolution procedure. The major contribution to the C 1s core level asymmetric peak is coming from the peak at 285 eV and we assigned it to graphitic C-C bonds [27]. The deconvolution of B 1s spectrum presented in Fig.8b showed two different chemical states at 189.5 and 192.5 eV. The peak at 282.8 eV in C 1s spectrum and 189.5 eV in B 1s spectrum corresponds to B-C bonds [28]. The deconvolution of the O 1s spectrum showed two different chemical states at 532.8 and 530.5 eV. Peak at 530.5 eV in the O 1s spectrum corresponds to native oxygen sitting on the sample surface [29]. The peak at 192.5 eV in B 1s spectrum and 532.8 eV in O 1s spectrum corresponds to boron oxide [30]. The surface of low thickness boron carbide like composed of mainly graphitic carbons and little amount of $B_4C$ and boron oxide like compounds.

### 4.5. TOF-SIMS measurements

Two boron carbide thin films of thickness ~9 and ~90 nm were re-deposited to check the repeatability of the measurements. The depth profiles of thin films of thicknesses 9 and 90 nm are plotted in Fig. 9(a) and (b) respectively, where, the X-axis represents time and the Y-axis the relative intensity. Assuming similar analysis conditions the relative intensities of $B^-$, $C_3^-$, $18O^-$ were compared to estimate their variation depending on deposited thickness. The ratio of B atom present in the 90 nm thin film to 9 nm thin film is ~2. This means that boron content increases with increasing the film thickness. The C and O content is high in 9 nm thin film as compared to 90 nm thin film. The higher O content in low thickness film is due to surface expose to ambient. TOF-SIMS confirms that the low thickness film contains more C and less B content as compared to high thickness films. The reason for such observed boron deficient or carbon rich growth in born carbide thin films is as follows.

With very few exceptions, evaporation and sputtering generates molecules of the starting compound or multi component material [31]. The atoms of components recombine on a surface (or in vapor phase if the number density of such atoms is large). Recombination on a surface depends on a large number of parameters. One such parameter is sticking coefficient which depends on substrate temp, substrate chemistry, chemical bonds between the arriving atoms, reactive gases present in the chamber, roughness and temperature of substrate, velocity and angle of vapor atoms on a substrate, surface impurities, etc. Sticking coefficient of B and C is expected to be different on a given substrate. In the initial stages depending on a whole range of parameters and this coefficient will increase with thickness.

## 5. Conclusions

Reflectivity measurements at different energies allow combining the sensitivity of GIXR data to microstructure parameters like layer thicknesses and interfacing roughness, with the layer composition sensitivity of RSXR. The optical index profile derived over extended region in the vicinity of the B K-edge was used to estimate the composition of the films quantitatively. At low thickness we observed boron deficient or carbon rich growth. TOF-SIMS data suggested that the decrease in boron content with decrease in film thickness. XPS measurements also suggested little amount of B atoms on the surface of low thickness sample. The RSXR technique is a photon in photon out processes which provides excellent spatial resolution, down ~ nm with penetration depths over hundreds of nanometers and gives the composition of films in nondestructive manner.

## Acknowledgement

The authors thank R. Dhawan for his help in sample preparation and Avinash wadikar for XPS measurements. Thanks are due to R. J. Choudhary and Madhusmita Baral for useful discussions. The authors are also thankful to Dr. B. Sundaravel for RRBS measurements and useful discussion. The authors are grateful to Prof. K. L. Chopra for useful discussion.

## List of figure captions

**Fig.1.** Measured and simulated GIXR curves of boron carbide thin films.

**Fig.2.** Measured and simulated RSXR curves of the sample S3 in the vicinity of B K-edge.

**Fig.3.** (a) Measured and simulated RSXR curves of the sample S2 in the vicinity of B K-edge, (b) in-depth optical index profile δ obtained from RSXR fits.

**Fig.4.** (color line) Optical constant profile of (a) lyer-2 and (b) layer-1 as a function of photon energy in the vicinity of B K-edge (black line) model composition consisting of different at.% B, C and O.

**Fig.5.** Measured and simulated RSXR curves of the sample S1 in the vicinity of B K-edge.

**Fig. 6**. RRBS spectra recorded with (a) 3.045 MeV alpha particles (O resonance), (b) 3.9 MeV alpha particles (B resonance) and (c) 4.27 MeV alpha particles (C resonance). The open circles are the experimental data and the solid line is the simulated data.

**Fig.7.** Over view XPS spectra of the samples S1 and S2.

**Fig.8.** High resolution XPS scans of the sample S1 in the vicinity of (a) C 1s, (b) B 1s and (c) O 1s lines together with spectrum deconvolution and Shirley background.

**Fig.9.** TOF-SIMS depth profile spectra of (a) 9 nm thin film (b) 90 nm thin film. SIMS signal of B, C, O and Si are shown in the graph.

**Table 1.** Thickness, density and roughness of deposited films obtained from the best fitted models shown in Fig. 1.

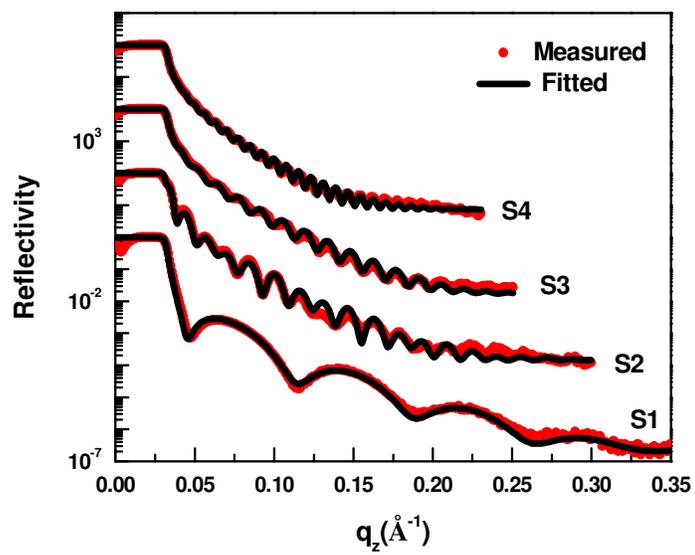

**Fig.1.** Measured and simulated GIXR curves of boron carbide thin films.

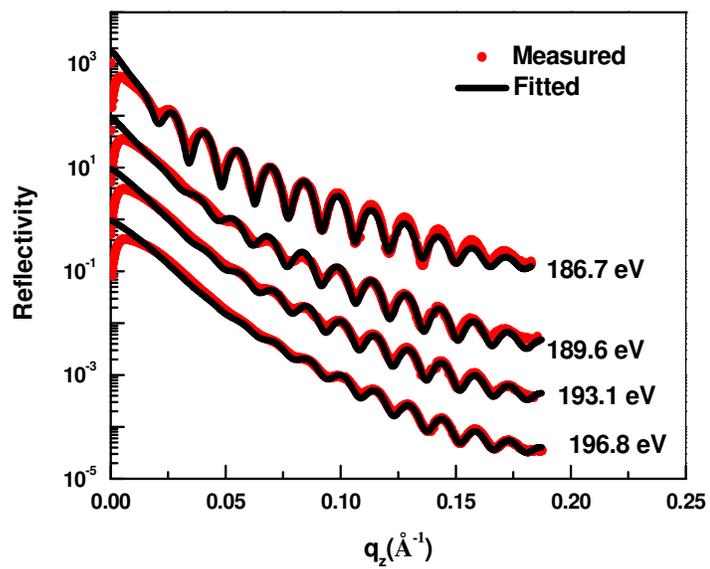

**Fig.2.** Measured and simulated RSXR curves of the sample S3 in the vicinity of B K-edge.

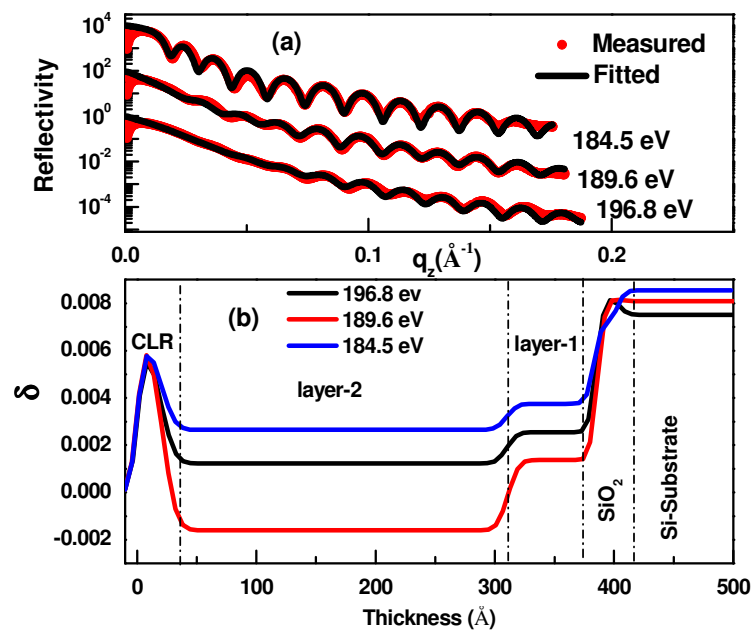

**Fig. 3.** (a) Measured and simulated RSXR curves of the sample S2 in the vicinity of B K-edge, (b) in-depth optical index profile δ obtained from RSXR fits.

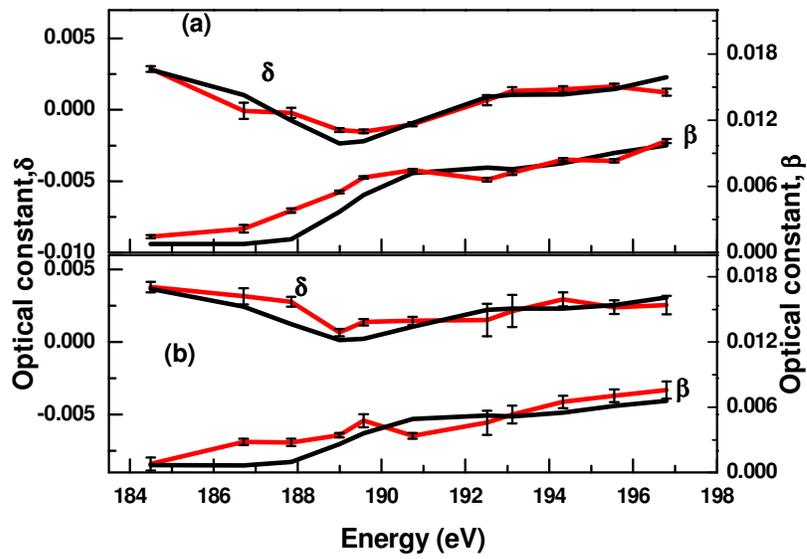

**Fig.4.** (color line) Optical constant profile of (a) lyer-2 and (b) layer-1 as a function of photon energy in the vicinity of B K-edge (black line) model composition consisting of different at.% B, C and O.

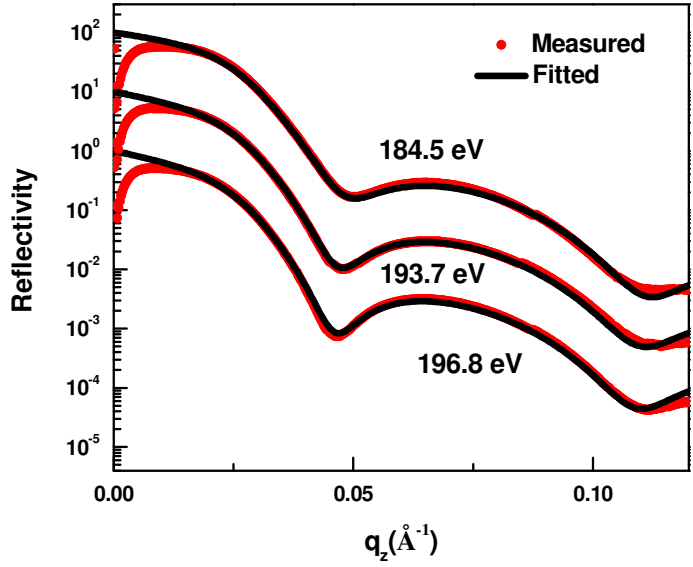

**Fig.5.** Measured and simulated RSXR curves of the sample S1 in the vicinity of B K-edge.

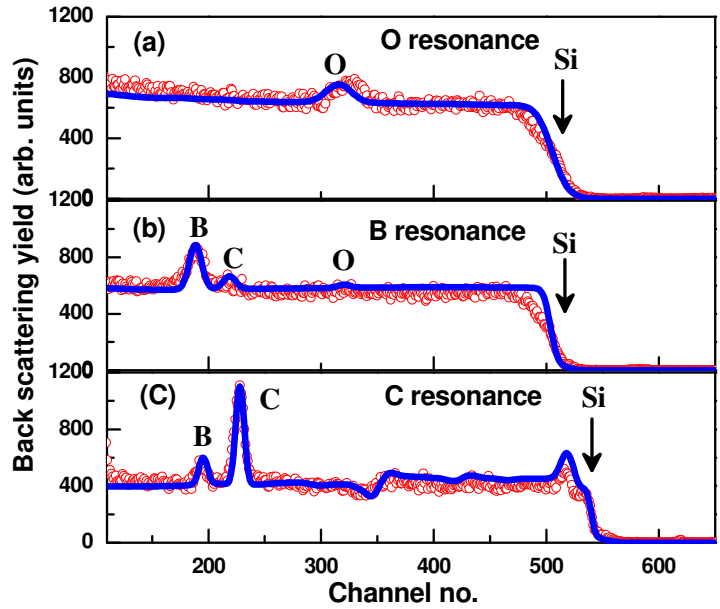

**Fig.6.** RRBS spectra recorded with (a) 3.045 MeV alpha particles (O resonance), (b) 3.9 MeV alpha particles (B resonance) and (c) 4.27 MeV alpha particles (C resonance). The open circles are the experimental data and the solid line is the simulated data.

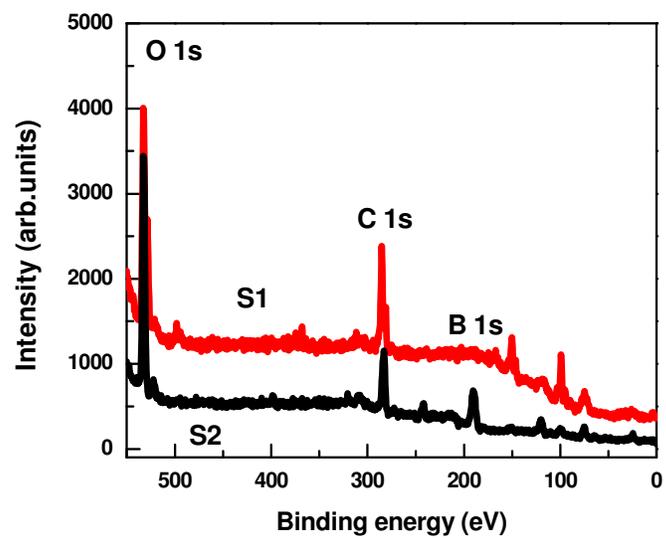

**Fig.7.** Overview XPS spectra of the samples S1 and S2.

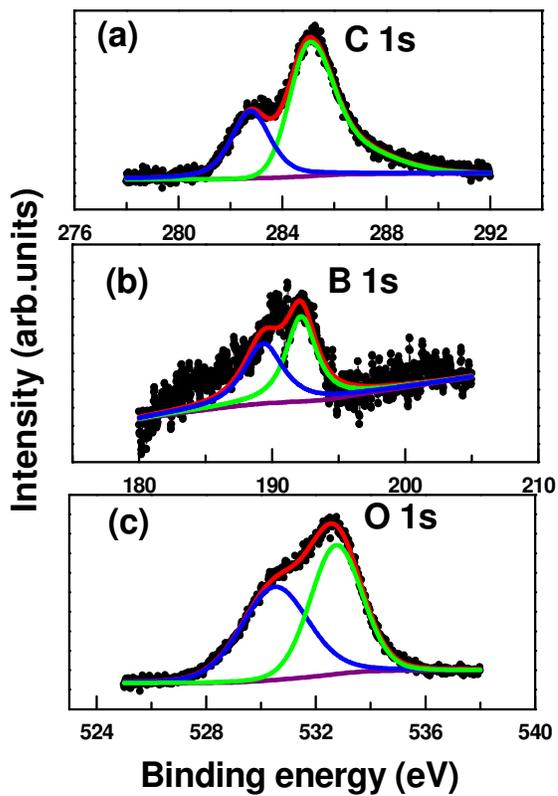

**Fig.8.** High resolution XPS scans of the sample S1 in the vicinity of (a) C 1s, (b) B 1s and (c) O 1s lines together with spectrum deconvolution and Shirley background.

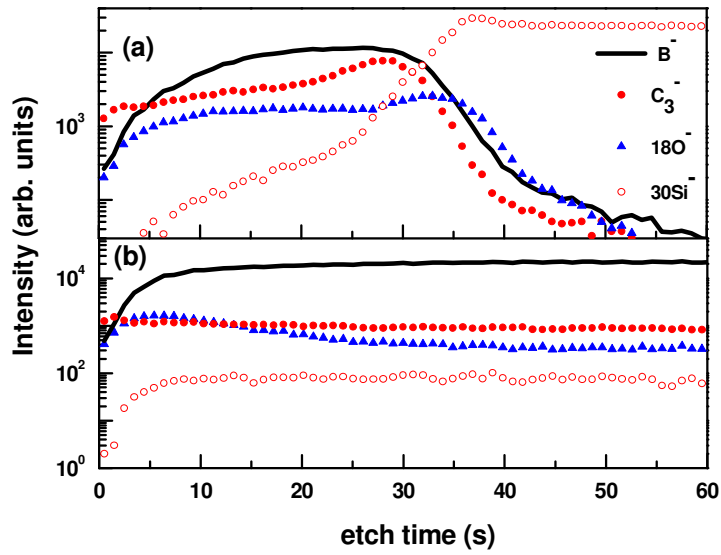

**Fig.9.** TOF-SIMS depth profile spectra of (a) 9 nm thin film (b) 90 nm thin film. SIMS signal of B, C, O and Si are shown in the graph.

**Table1.** Thickness, density and roughness of deposited films obtained from the best fitted models shown in Fig.1.

| Sample name | Thickness (nm) | Density (g/c.c) | Roughness (nm) |
|---|---|---|---|
| S1 | 8.35±0.05 | 1.76±0.07 | 0.6 |
| S2 | 29.3±0.2 | 2.46±0.07 | 0.9 |
|    | 7.2±0.2 | 2.08±0.07 | 0.7 |
| S3 | 40.8±0.2 | 2.50±0.04 | 1.0 |
| S4 | 76.8±0.3 | 2.50±0.04 | 1.2 |